\documentstyle[a4,11pt,epsf]{article}
\begin{document}

\vspace*{10mm}

\begin{center}  
\sf
{\LARGE \bf INSTITUT F\"UR HOCHENERGIEPHYSIK \\}

\vspace*{1cm}
\rm
\hspace{120mm}
\parbox[t]{40mm}{
{\large \tt PHE 90-30 
\\
     LMU 90-3
\\
     November 1990
\\
     hep-ph/9808372}
}

\vspace*{20mm}
{\LARGE \bf 
                    Radiative Corrections to Deep Inelastic Scattering in the 
                    Presence of an Additional $Z'$ at $LEP \times LHC$ 
} \\
\vspace*{25mm}
{\LARGE \tt
                   J.~Bl\"umlein
\\ \bigskip
 Institut f\"ur Hochenergiephysik, Zeuthen
\\ \bigskip
  and
\\ \bigskip
  Sektion Physik der Universit\"at M\"unchen
\\ \bigskip
A.~Leike, T. Riemann
\\ \bigskip
                    Institut f\"ur Hochenergiephysik, Zeuthen
} 

\vspace*{30mm}
{\Large \it Contribution to the Proc. of the ECFA Large Hadron
Collider Workshop, Aachen, 4-9 Oct 1990, CERN 90-10 } 

\vspace*{15mm}

\sf
{\LARGE \bf PLATANENALLEE 6 $\cdot$ O-1615 ZEUTHEN} \\

\vspace{3mm}
{\LARGE \bf GERMANY}
\end{center}

\setcounter{page}{0}
\clearpage
%
%%%%%%%%%%%%%98\end{document}
%=====================================================================
%  lhcproc.tex follows:
%98\documentstyle[11pt,a4]{article}
%               30.11.90 -- 1
% formal changes July 1998
\title{
%\vspace{-1.4cm} 
%\begin{flushright}                
%{\large PHE 90-30} 
%\end{flushright}
%\vspace{.5cm} 
\LARGE 
{\bf
                 Radiative Corrections to Deep Inelastic Scattering in 
\\
the 
                 Presence of an Additional $Z'$ at $LEP \times LHC$}
            %     \thanks{Contribution to the ECFA Large Hadron Collider 
            %             Workshop, Aachen, 4-9 Oct 1990, to appear in the 
            %             proceedings}
}
\author
{
                            J. Bl\"umlein, A. Leike, T. Riemann\\
                            Institut f\"ur Hochenergiephysik, Platanenallee 6,
                            O-1615 Zeuthen,Germany
}
\date{}
%                             11 November 1990
%} 
\newcommand{\leplhc}{$LEP \times LHC$ \,}
\newcommand{\ee}{$e^+e^-\ $}
\newcommand{\afb}{$A_{FB}\:$}
\newcommand{\st}{$\sigma_T\:$}
\newcommand{\oalf}{O($\alpha$)$\:$}
\newcommand{\ff}{$f^+f^-\ $}
\newcommand{\mumu}{$\mu^+ \mu^-\:$} 
\newcommand{\nobody}{\rule{0ex}{1ex}}
 
%%%%%%%%%98\arabic{section}
%\renewcommand{\baselinestretch}{1.5}
 
%%%%%%%98\begin{document}

\maketitle
%*********************************************
\begin{abstract}
The study of ep-scattering at \leplhc offers an interesting 
opportunity to search for additional heavy neutral gauge bosons. We
study the influence of radiative corrections, especially of the 
potentially large QED contributions, to the expected cross sections.
These corrections are typically of the same order as the effect searched
for or even larger and have to be taken into account properly.
\end{abstract}
%\vfill\eject
%********************************************
 
\section{Formulation of the Problem}
Neutral current scattering of electrons and protons proceeds via the 
exchange of photons and $Z$ bosons. If there exist  additional heavy gauge 
bosons in Nature, these would also give some contribution to the cross 
section. So, they could be detectable at the \leplhc. The experimental 
sensitivity of the reaction 
%---------------
\begin{equation}
e^-p \longrightarrow (\gamma,Z,Z') \longrightarrow e^-X
\label{eq1}
\end{equation}
%--------------
to details of the $Z'$ model has been studied earlier (~$^{\cite{oldref}}$ 
and refs. quoted therein) and also in a contribution to this 
workshop $^{\cite{martyn}}$. It is well-known that the Born cross section may 
be modified strongly by radiative corrections of the standard model. 
These effects have to be under definite control. For the annihilation channel,
the problem has been solved in $^{\cite{leiri}}$ where some of the notations 
used may be found. Here, we calculate the weak and 
QED corrections to reaction (\ref{eq1}) in presence of an additional heavy gauge 
boson $Z'$.

The Born cross section may be denoted as follows:
%------------------------
\begin{equation}
\frac{d\sigma_{0}}{dxdy} = 2\pi \alpha^{2} \frac{s}{Q^{4}} 
\left[ \sigma(\gamma,\gamma) +2\sigma(\gamma,Z) +\sigma(Z,Z) +2\sigma(
\gamma,Z') 
      +2\sigma(Z,Z')          +\sigma(Z',Z') 
\right]. 
\label{eq2}
\end{equation}
%------------------------
In a compact notation, the six contributions are:
%------------------------
\begin{equation}
\sigma(B_{i},B_{j}) = \chi_{i}(Q^{2}) \chi_{j}(Q^{2})
    \left[ Y_{+} {\bf V} +  Y_{-} {\bf A} \right], 
\end{equation}
%------------------------
\begin{equation}
\chi_{n}(Q^{2}) = \frac{g_{n}^{2}}{4\pi\alpha} \frac{Q^{2}}{Q^{2}+M_{n}^{2}},
\hspace{1.cm}
Y_{\pm} =  \left[ 1 \pm (1-y) \right]^{2},
\end{equation}
%------------------------
\begin{equation}
{\bf V} = \left[ C_{V}(e) + \lambda Q_{e} C_{A}(e) \right] \times
          \left[ C_{V}(u) (u + \bar u) + C_{V}(d) (d + \bar d) \right],
\end{equation}
%------------------------
\begin{equation}
{\bf A} =- \left[ Q_{e} C_{A}(e) + \lambda C_{V}(e) \right] \times
           \left[ C_{A}(u) (u - \bar u) + C_{A}(d) (d - \bar d) \right],
\end{equation}
%------------------------
\begin{eqnarray}
C_{V}(f) = v_{f}(i) v_{f}(j) + a_{f}(i) a_{f}(j),
\hspace{1.cm}
C_{A}(f) = v_{f}(i) a_{f}(j) + a_{f}(i) v_{f}(j).
\end{eqnarray}
%------------------------
%%%\begin{equation}
%C_{A}(f) = v_{f}(i) a_{f}(j) + a_{f}(i) v_{f}(j).
%\end{equation}
%-------------------------
The couplings in the standard theory are:
%------------------------
\begin{equation}
g_{0}=e, \; g_{1}=\sqrt{\sqrt{2}G_{\mu}M_{Z}^{2}}, \; v_{f}(0)=Q_{f}, \; 
a_{f}(0)=0, \; v_{f}(1)=I_{3}^{L}-2Q_{f}\sin^{2}\theta_{W}, \; 
a_{f}(1)=I_{3}^{L}.
\end{equation}
%------------------------
Here, $i=0,1$ are photon and $Z$ boson with $Q_{e}=-1,I_{3}^{L}(e)=-1/2$.

We now discuss the inclusion of the QED corrections. For the standard theory,
for the $\gamma Z'$ interference, and for the pure $Z'$ contribution,  
this may be done following e.g. $^{\cite{bbcr}}$ correct to order $O(\alpha)$ 
or e.g. $^{\cite{blum}}$ in the leading log approximation (LLA). The term to be
determined additionally
is the $ZZ'$ interference. 
We have calculated the bremsstrahlung corrections in LLA 
which is known to agree with the exact calculation within about 
1~\% or better. We took into account all the contributions which are discussed
in~$^{\cite{blum}}$ where also the notational details may be found. The general
form of the bremsstrahlung terms is:
%------------------------
\begin{equation}
d\sigma_{QED} = \frac{\alpha}{2\pi} \ln \frac{Q^{2}}{m_{e}^{2}}
\int_{0}^{1} dz \frac{1+z^{2}}{1-z} 
\left\{ \theta(z-z_{0} ) \frac{y}{\hat y} \frac{1}{z^{a}} 
%A%\{ \theta(z-z_{0} ) \frac{y}{\hat y} \frac{1}{z^{a}} 
\frac  {d\sigma_{0}}  {dxdy} |_{x=\hat x, y=\hat y, s=\hat s}
%A%\frac  {d\sigma_{0}}  {dxdy} (Z,Z') {
%A%\left| \nobody_{ \small \begin{array}{l}
%A%x=\hat x  \\
%A%y=\hat y  \\
%A%s=\hat s
%A%\end{array} } \right. }
%A%-  \frac  {d\sigma_{0}}  {dxdy} (Z,Z') \} .
-  \frac  {d\sigma_{0}}  {dxdy}  \right\} .
\end{equation}
%------------------------

Finally, the weak loop corrections of the standard theory are taken into 
account using the form factor approach~$^{\cite{bbcr}}$. This leads to an
introduction of finite renormalisation factors for the Fermi constant from 
muon decay and for the weak mixing angle as being defined in the on mass shell 
renormalisation scheme:
%------------------------
\begin{equation}
G_{\mu} \; \rightarrow \; G_{\mu} \, \rho(s,Q^{2},\alpha,M_{Z},M_{H},m_{t},
\ldots),  \hspace{1cm}
\sin^{2}\theta_{W} \; \rightarrow \; \sin^{2}\theta_{W} \, 
\kappa(s,Q^{2},\alpha,M_{Z},M_{H},m_{t},\ldots).
\end{equation}
%------------------------
In the effective Born cross section, we also use a running 
$\alpha_{QED}(Q^{2})$.
Leaving out the complex details of weak loop effects, we quote only
the  leading, universal t-quark mass effects~$^{\cite{bhr}}$:
%------------------------
\begin{equation}
\rho \; \approx \; \frac{1}{1-\delta \bar\rho},
\hspace{1cm}
\kappa \; \approx \; 1 + \frac {\cos^{2}\theta_{W}}  {\sin^{2}\theta_{W}} 
\delta \bar\rho,
 \hspace{1cm}
\delta \bar\rho = 3 \frac {G_{\mu}} {\sqrt{2}} \frac {m_{t}^{2}} {8\pi^{2}} 
\left[ 1 + \frac {G_{\mu}} {\sqrt{2}} \frac {m_{t}^{2}} {8\pi^{2}} 
\left( 19 - 2 \pi^{2} \right) \right].
\end{equation}
%------------------------
%----------------------------------------------------------------------
\section{Extra Z bosons}
 
   We study the interaction of photon, standard $Z$
boson and one extra $Z$ boson~$Z'$ with fermions:
\begin{equation}
{\cal L} = eA_\mu J_\gamma^\mu + g_1 Z_\mu J_Z^\mu + g_2 Z'_\mu J_{Z'}^\mu,
\label{eq31}
\end{equation}
where the currents are of the form
\begin{equation}
J_n^\mu = \sum_{f}\;\bar{f} \gamma^\mu\;[v_f(n) + \gamma_5 a_f(n)]\; f,\ \
n = \gamma, Z, Z'.
\label{eq32}
\end{equation}
The physical mass eigenstates $Z_1$  and $Z_2$
are the result of a mixing between $Z$ and $Z'$. 
From experimental data the
corresponding mixing angle is known to be very small,
$\cos \theta_{M} \approx 0$.
 
   We now investigate numerical consequences of extra $Z$ bosons arising 
from a grand unified theory based on the
$E_6$  group$^{\cite{ze6}}$:
\begin{eqnarray}
E_6 \rightarrow \ldots
%%%SO(10) \times U(1)_\psi \rightarrow
%%%SU(5) \times U(1)_\chi \times U(1)_\psi \rightarrow 
%%%\nonumber \\ 
\rightarrow
SU(3)_c \times SU(2)_L \times U(1)_Y \times U(1)_\chi \times U(1)_\psi .
%\label{eq33}
\end{eqnarray}
 
   The following linear combination of
the two extra $Z$ bosons $Z_\chi$ of $U(1)_\chi$ and $Z_\psi$ of $U(1)_\psi$ is
assumed to be light$^{\cite{zmix}}$:
\begin{equation}
Z' =\cos \theta_E\;Z_\chi + \sin \theta_E\;Z_\psi.
\label{eq34}
\end{equation}
  We consider three models which differ in their phenomenological
consequences considerably as has been studied in$^{\cite{martyn}}$: 
model A (B, C) with $\theta_{E} = -0.9117$ \, (0., 1.3181). The relation 
to the mixing angle $\alpha$ of ref.$^{\cite{martyn}}$ is:
$\theta_{E} = \alpha -\pi/2$. 
%---------------------------------------------------------------------------
\section{Results}

   We discuss radiative corrections in presence of a $Z'$ for parameters and 
observables as used in $^{\cite{martyn}}$:
\begin{equation}
A_{LR}^{--} = \frac { d\sigma(e_{L}^{-}) - d\sigma(e_{R}^{-}) }
                     { d\sigma(e_{L}^{-})  + d\sigma(e_{L}^{-}) },
\hspace{1.cm}
A_{LL}^{-+} = \frac { d\sigma(e_{L}^{-}) - d\sigma(e_{L}^{+}) }
                     { d\sigma(e_{L}^{-})  + d\sigma(e_{L}^{+}) },
\hspace{1.cm}
A_{RR}^{-+} = \frac { d\sigma(e_{R}^{-}) - d\sigma(e_{R}^{+}) }
                     { d\sigma(e_{R}^{-})  + d\sigma(e_{R}^{+}) },
\end{equation}
where the degree of polarisation has been chosen to be $\lambda = 0.8$.

 At small $y$, where also the energy scale set by $Q^2$ is 
small, both the left-right asymmetries and the corrections to them are small 
since the symmetric pure photon exchange dominates. Although the largest 
cross section corrections are due to bremsstrahlung (especially at large $y$),
being followed by those due to the running $\alpha$, this is not true for
asymmetries as is seen best from figs. 1 and 2. 
There, the loop corrections 
dominate over bremsstrahlung. Further, a closer inspection shows that the 
influences from a 
running $\alpha$ and the weak form factor effects are of the same size 
and have equal sign. 
   For comparison, the curves from the standard model predictions are also 
shown. They are
corrected by LLA bremsstrahlung, by
effects due to running $\alpha_{QED}$ and the weak form factors.   
As may be seen they behave such that a distinct experimental signal
from a $Z'$ deserves a careful analysis of data.

   We conclude that in the analysis of experimental 
asymmetries all different origins of radiative corrections have to be
carefully taken into account. This is even more important for the charge
asymmetries as shown in figs. 3 and 4.

%***********************************************

%\end{document}
%######################################################################
% follows file figs.tex from AL.
%\documentstyle[11pt,epsf]{article}
%\setlength{\textwidth}{168mm}
%\setlength{\textheight}{230mm}
%\begin{document}
\thispagestyle{empty}
\begin{figure}[tbh]
%-----------Fig.1-----------------------------
\ \vspace{1cm}\\
\begin{minipage}[t]{7.8cm} {
\begin{center}
\hspace{-1.7cm}
\mbox{
\epsfysize=7.0cm
\epsffile[0 0 500 500]{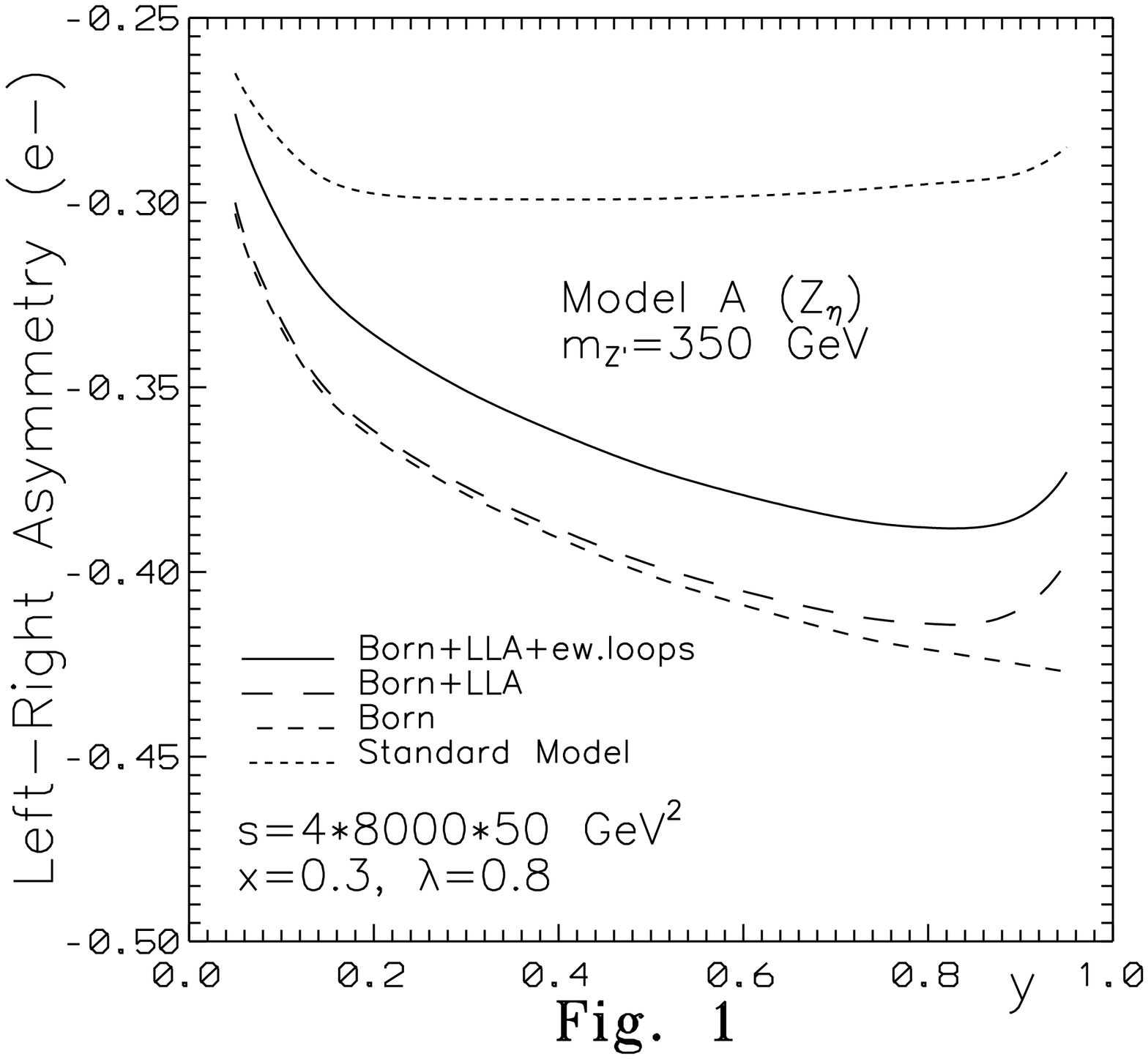}
}
\end{center}
%\vspace*{-0.5cm}
%\noindent
%{\small\it {\bf Figure 1\ \ }
%Figure caption 
%}
}\end{minipage}
\hspace*{0.5cm}
% ----------------Fig. 2-----------------------
\begin{minipage}[t]{7.8cm} {
\begin{center}
\hspace{-1.7cm}
\mbox{
\epsfysize=7.0cm
\epsffile[0 0 500 500]{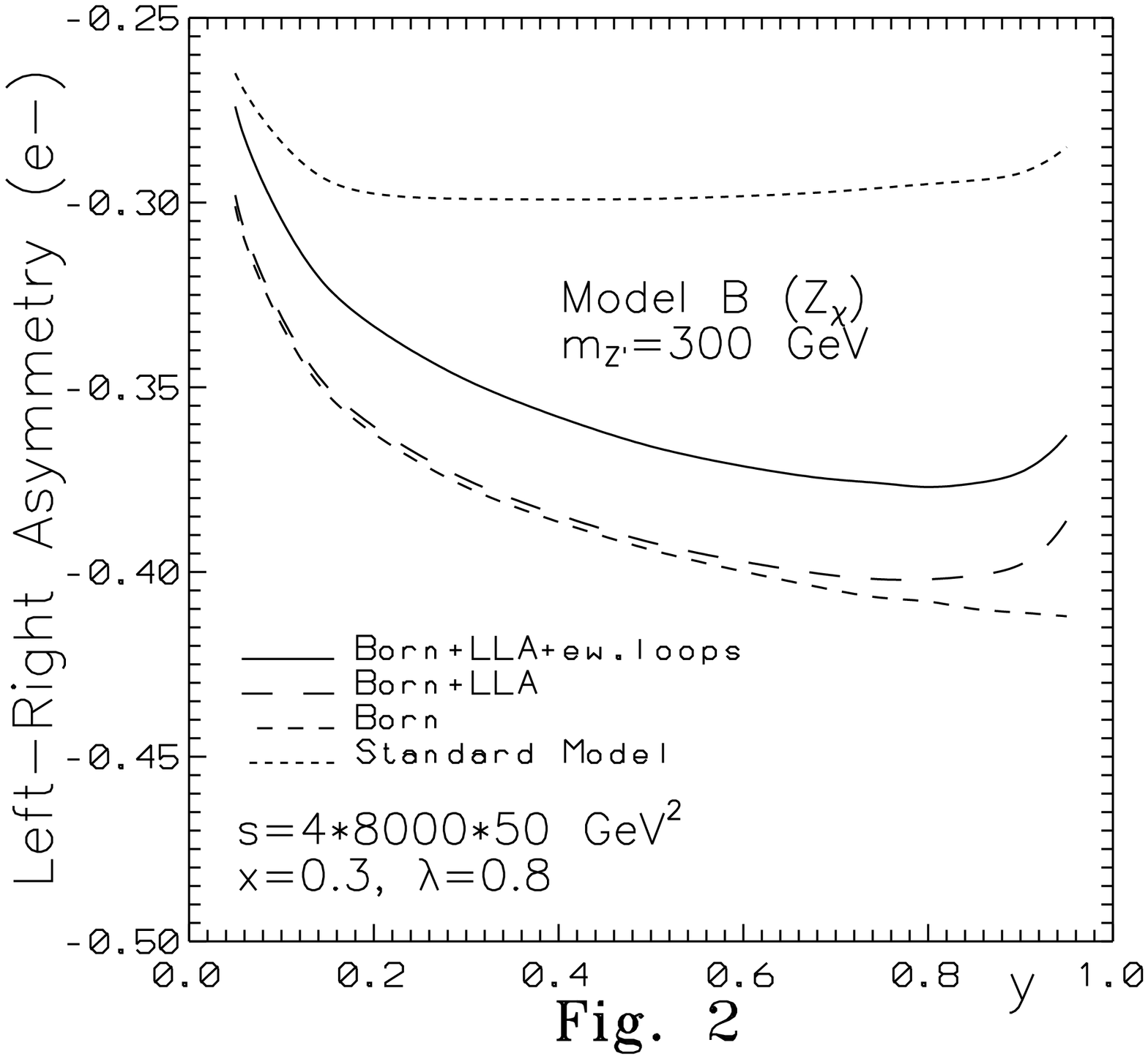}
}
\end{center}
%\vspace*{-0.5cm}
%\noindent
%{\small\it {\bf Figure 2\ \ }
%Figure caption 
%}
}\end{minipage}
\end{figure}

\begin{figure}[tbh]
%-----------Fig.3-----------------------------
\ \vspace{1cm}\\
\begin{minipage}[t]{7.8cm} {
\begin{center}
\hspace{-1.7cm}
\mbox{
\epsfysize=7.0cm
\epsffile[0 0 500 500]{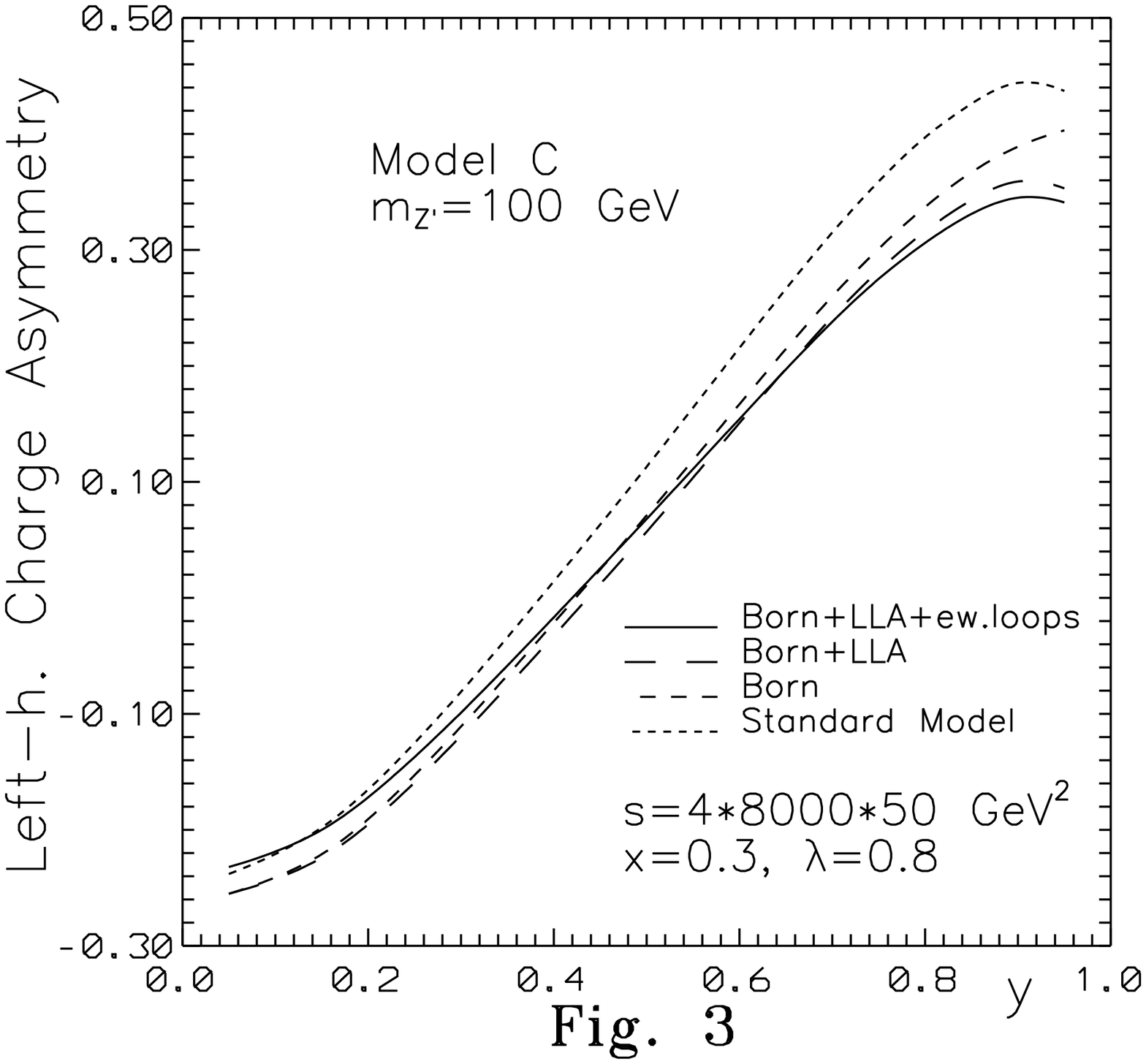}%
}
\end{center}
%\vspace*{-0.5cm}
%\noindent
%{\small\it {\bf Figure 3\ \ }
%Figure caption 
%}
}\end{minipage}
\hspace*{0.5cm}
% ----------------Fig. 4-----------------------
\begin{minipage}[t]{7.8cm} {
\begin{center}
\hspace{-1.7cm}
\mbox{
\epsfysize=7.0cm
\epsffile[0 0 500 500]{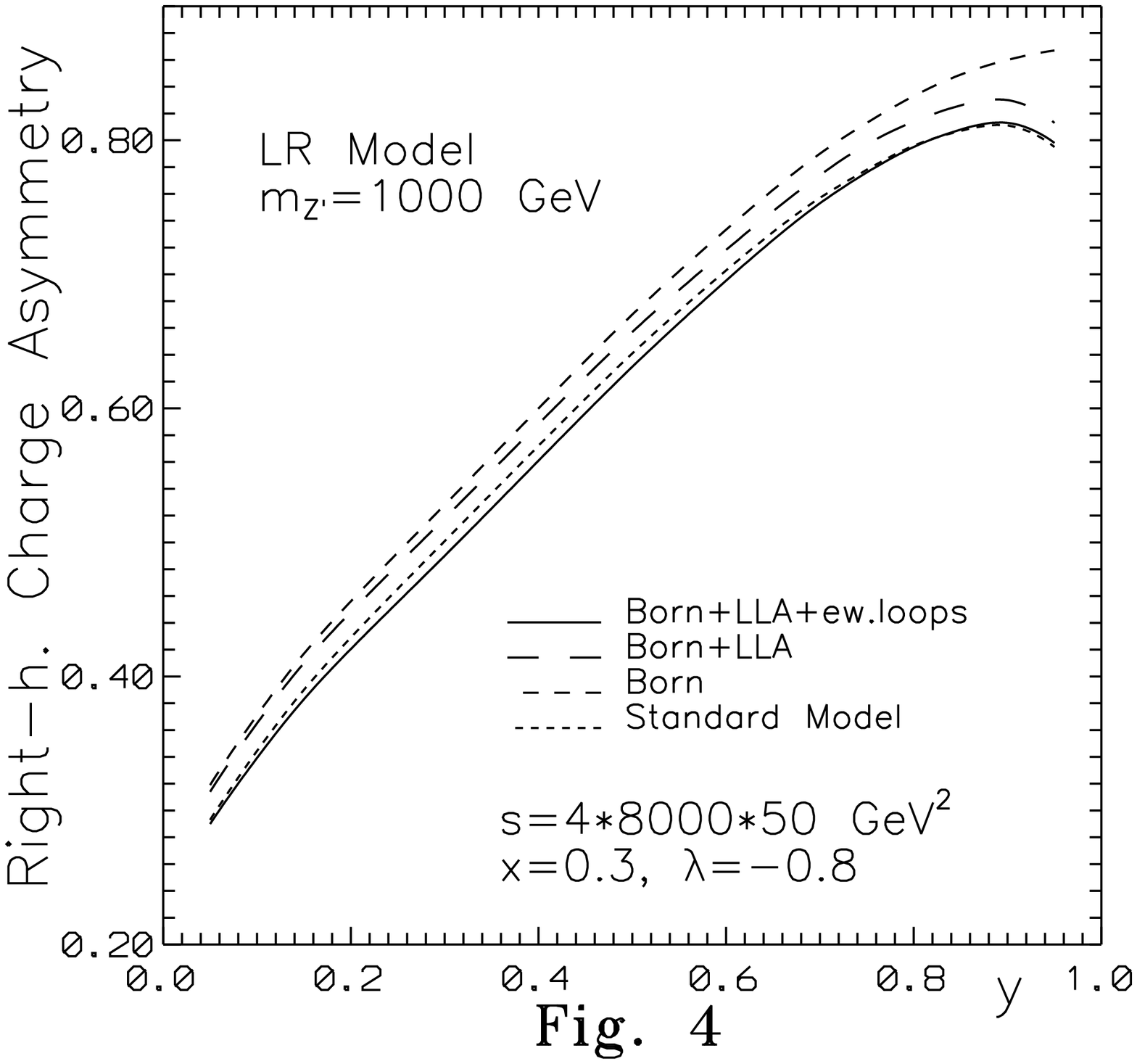}
}
\end{center}
%\vspace*{-0.5cm}
%\noindent
%{\small\it {\bf Figure 4\ \ }
%Figure caption 
%}
}\end{minipage}
\end{figure}
\end{document}